\documentstyle[12pt,epsfig]{article}

\newcommand{\be}{\begin{equation}}
\newcommand{\ee}{\end{equation}}

\newcommand{\dlt}{\delta}

\newcommand{\br}{{\bf r}}
\newcommand{\bS}{{\bf S}}

\newcommand{\bB}{{\bf B}}

\newcommand{\bt}{\beta}

\newcommand{\al}{\alpha}

\newcommand{\gm}{\gamma}
\newcommand{\om}{\omega}

\newcommand{\Gm}{\Gamma}

\newcommand{\lbd}{\lambda}

\begin{document}

\begin{center}
{\Large{\bf Coherent spin radiation by magnetic nanomolecules
and nanoclusters} \\ [5mm]
V.I. Yukalov$^1$, V.K. Henner$^{2,3}$, P.V. Kharebov$^3$, and
E.P. Yukalova$^4$} \\ [3mm]

{\it
$^1$Bogolubov Laboratory of Theoretical Physics, \\
Joint Institute for Nuclear Research, Dubna 141980, Russia\\ [3mm]
$^2$University of Louisville, Louisville, Kentucky 40292, USA \\ [3mm]
$^3$Perm State University, Perm 614000, Russia \\ [3mm]
$^4$Laboratory of Information Technologies, \\
Joint Institute for Nuclear Research, Dubna 141980, Russia
}

\end{center}

\vskip 2cm

\begin{abstract}

The peculiarities of coherent spin radiation by magnetic nanomolecules
is investigated by means of numerical simulation. The consideration is
based on a microscopic Hamiltonian taking into account realistic dipole
interactions. Superradiance can be realized only when the molecular
sample is coupled to a resonant electric circuit. The feedback mechanism
allows for the achievement of a fast spin reversal time and large radiation
intensity. The influence on the level of radiation, caused by sample
shape and orientation, is analysed. The most powerful coherent radiation
is found to occur for an elongated sample directed along the resonator
magnetic field.

\end{abstract}

\vskip 1cm
{\parindent=0pt
{\bf Key words}: coherent radiation; radiowave sources; molecular magnets;
relaxation effects; numerical simulation; magnetic anisotropy

\vskip 1cm

 }

\newpage

\section{Nanomolecules and nanoclusters}

The possibility of realizing coherent spin relaxation, in the regime
of {\it collective induction}, was advanced by Bloembergen and Pound [1],
who suggested to couple the spin sample to a resonant electric circuit.
In the presence of a resonator, there can also arise five other regimes
of coherent spin motion producing coherent radiation: {\it maser generation,
pure superradiance, triggered superradiance, pulsing superradiance, and
punctuated superradiance}, whose detailed description can be found in
review articles [2,3]. At the first stage, coherent relaxation effects
have been studied for nuclear spins [4--10]. Recently [11--14], the
possibility of coherent spin relaxation in molecular magnets, possessing
large electronic spins, was suggested.

Molecular magnets are composed of magnetic nanomolecules having several
properties essentially distinguishing molecular spins from nuclear spins
(see review articles [3,11,15,16]). First of all, magnetic molecules can
possess different values of the total spin, including rather large spins.
As examples we can mention the following molecules, whose spins are shown
in brackets:
$$
{\rm
K_6 \left [ V_{15}^{4+} As_6 \; O_{42} (H_2O) \right ] \; 8H_2 O
\qquad \left ( S = \frac{1}{2}\right ) \; ,}
$$
$$
{\rm
(Ph\; Si\; O_2)_6 Cu_6 (O_2\; Si\; Ph)_6 \qquad (S=3) \; , }
$$
$$
{\rm
\left [ Mn_{12} O_{12} (CH_3COO)_{16} (H_2O)_4\right ]
(2CH_3COOH)\; 4H_2O \qquad (S=10) \; , }
$$
$$
{\rm
\left [ Fe_8 O_2 (OH)_{12} T_6 \right ]^{8+} \qquad (S=10) \; , }
$$
$$
{\rm
Mn_6 O_4 Br_4 (Et_2\; dbm)_6  \qquad (S=12) \; , }
$$
$$
{\rm
\left [ Cr (C N\; Mn\; L)_6  \right ](Cl\; O_4)_9 \qquad
\left ( S = \frac{27}{2} \right ) \; . }
$$

The linear size of such molecules is about 1nm$=10\AA$. A system of these
molecules forms a crystalline cluster with a well organized crystalline
lattice, whose spacing is around $14\AA$. Below the blocking temperature
of approximately $1$K, each magnetic nanomolecule possesses, in its ground
state, a total spin $S$, that is, the magnetic moment $\mu_BS$. The
magnetization reversal is caused by the weak spin-phonon interactions,
because of which the reversal process is extremely slow, lasting for
$T_1\sim 10^5-10^7$ s. More detailed information on magnetic nanomolecules
can be found in Refs. [3,11,15,16].

In addition to magnetic nanomolecules, there exists a whole series of
magnetic nanoclusters that also exhibit large total spins [11,17--19].
Such clusters can be formed by metals, e.g., Ni, Fe, Co, Hg, by oxides,
NiO, Fe$_2$O$_3$ and like that, and by alloys, for instance, NiFe$_2$O$_4$,
Nd$_2$Fe$_{14}$B, Pr$_2$Fe$_{14}$B, Tb$_2$Fe$_{14}$B, DyFe$_{14}$B,
Pr$_2$Co$_{14}$B, Sm$_1$Fe$_{11}$Ti$_1$, Sm$_1$Fe$_{10}$V$_2$,
Sm$_2$Fe$_{17}$N$_{23}$, Sm$_2$Fe$_{17}$C$_{22}$, Sm$_2$Co$_{17}$, and
SmCo$_5$. These nanoclusters can be of various sizes, with diameters
between $10\AA$ to $10^4\AA$. The total spin $S$ can range between $10^2$
to $10^4$. So, the total magnetic moment can be very large. For illustration,
we may give some parameters for iron-platinum nanoclusters [20-22]. Thus,
the nanocluster compound Fe$_{50}$Pt$_{50}$ possesses an effective spin
$S\sim 2\times 10^4$. The Curie temperature plays the role of the blocking
temperature, for this material being $T_c=710$ K. The uniaxial anisotropy
parameter $D=K_cV_c$, with the anisotropy strength $K_c=0.6\times 10^8$
erg/cm$^3$. The typical diameter of $63\AA=6.3$ nm corresponds to a spherical
cluster of volume $V_c=1.31\times 10^{-19}$ cm$^3$. So, the anisotropy
parameter is $D=0.785\times 10^{-11}$ erg, and $D/\hbar=0.745\times 10^{16}$
s$^{-1}$. The escape time for temperatures below $T_c$ is larger than
$10^{32}$ s. The nanoclusters of Fe$_{70}$Pt$_{30}$ have effective spins
$S\sim 2\times 10^3$, Curie temperature $T_c=420$ K, and the cubic anisotropy
strength $K_c=0.8\times 10^7$ egr/cm$^3$. With a typical diameter of
$23\AA=2.3$ nm, the volume of a spherical cluster is $V_c=0.637\times
10^{-20}$ cm$^3$.  The corresponding anisotropy parameter $D=0.51\times
10^{-13}$ erg, and $D/\hbar=0.483\times 10^{14}$ s$^{-1}$. Escape times at
low temperatures are extremely long.

The problem, we address in the present paper, is as follows. Suppose we
consider a system of nanomolecules or nanoclusters. For short, we shall
mention in what follows nanomolecules, keeping in mind that the same
consideration is applicable to nanoclusters. Being placed in an external
magnetic field, the system can be strongly spin polarized. Then, the system,
magnetized in one direction, is placed in an external magnetic field of the
opposite direction. That is the way of preparing a strongly nonequilibrium
system. The natural process of magnetization reversal in such a system is
extremely slow, being characterized by the time $T_1$. The reversal can be
made essentially faster by applying strong transverse external fields, either
static or alternating [3,19,23]. However, the fastest relaxation can be
achieved by coupling the spin sample with a resonator [3,11--14], when the
relaxation process becomes coherent.

The faster the relaxation process, the stronger the radiation intensity
produced by moving spins. And the coherent spin relaxation results in
coherent spin radiation. As has been mentioned above, there are several
types of coherent spin radiation [2,3]. The fastest spin relaxation and,
respectively, the strongest radiation intensity occurs, when the process
is self-organized. Such a self-organized coherent radiation is called {\it
superradiance}.

It is important to emphasize that the conditions of spin superradiance
are very different from those of atomic superradiance [24,25] and
acoustic superradiance [26]. In spin systems, because of the strong
decoherence,  caused by dipole interactions, superradiance is impossible
without a resonator [3,11--14,27].

The aim of the present paper is to find the optimal conditions under which
spin superradiance reaches the largest radiation intensity.

\section{Equations of motion}

The consideration is based on the microscopic Hamiltonian
\be
\label{1}
\hat H = \sum_i \hat H_i \;  + \; \frac{1}{2} \;
\sum_{i\neq j} \hat H_{ij} \; ,
\ee
in which $\hat H_i$ is a single-molecule term and $\hat H_{ij}$ describes
molecular spin interactions. The single-particle term
\be
\label{2}
\hat H_i = -\mu_0 \bB \cdot \bS_i - D(S_i^z)^2
\ee
includes the Zeeman energy, with the electron magnetic moment $\mu_0=-2\mu_B$,
and the single-site magnetic axial anisotropy term, with the anisotropy parameter
$D$. Generally, there can also be present a cubic anisotropy, with a term of the
type $-D[(S_i^x)^2(S_i^y)^2+(S_i^y)^2(S_i^z)^2+(S_i^z)^2(S_i^x)^2]$. However,
such terms are usually small. The magnetic field
\be
\label{3}
\bB = B_0 {\bf e}_z + H {\bf e}_x
\ee
contains an external static magnetic field $B_0$ and the resonator feedback
field $H$. The interaction term
\be
\label{4}
\hat H_{ij} = \sum_{\al\bt} D_{ij}^{\al\bt} \; S_i^\al\; S_j^\bt
\ee
is due to dipole interactions, with the dipolar tensor
\be
\label{5}
D_{ij}^{\al\bt} = \frac{\mu_0^2}{r_{ij}^3} \;
\left ( \dlt_{\al\bt} - 3 n_{ij}^\al\; n_{ij}^\bt \right ) \; ,
\ee
where
$$
r_{ij} \equiv |\br_{ij}| \; , \qquad {\bf n}_{ij} \equiv
\frac{\br_{ij}}{r_{ij}} \; , \qquad \br_{ij} \equiv \br_i -\br _j \; .
$$
The resonator is an electric circuit surrounding the spin sample [3]. The
resonator feedback field is formed by a magnetic coil and is described by
the Kirchhoff equation
\be
\label{6}
\frac{dH}{dt} + 2\gm H + \om^2 \int_0^t \; H(t')\; dt' =  - 4\pi\eta\;
\frac{dm_x}{dt} \; ,
\ee
in which $\gm$ is the resonator damping, $\om$ is the resonator natural
frequency, $\eta$ is a filling factor, and
\be
\label{7}
m_x \equiv \frac{\mu_0}{V} \; \sum_j < S_j^x >
\ee
is the transverse magnetization density, $V$ being the sample volume.
Molecules are enumerated by the index $j=1,2,\ldots,N$, each molecule
possessing spin $S$, with the spin vector $\bS_i=\{ S_i^\al\}$. In what
follows, we set the filling factor $\eta=1$.

The characteristic frequencies of the system are the Zeeman frequency
\be
\label{8}
\om_0 = -\; \frac{\mu_0}{\hbar} \; B_0 =  \frac{2}{\hbar} \; \mu_B B_0 \; ,
\ee
the resonator natural frequency $\om$, and the anisotropy frequency
\be
\label{9}
\om_D = (2S - 1) \frac{D}{\hbar} \; .
\ee
The resonance conditions
\be
\label{10}
\left | \frac{\om -\om_0}{\om_0} \right | \ll 1 \; , \qquad
\frac{\om_D}{\om_0} \ll 1
\ee
are assumed. For typical magnetic molecules, $\om_D\sim(10^{10}-10^{12})$
s$^{-1}$.

The most important relaxation parameters are the spin-phonon attenuation
$\gm_1\equiv 1/T_1$, the spin-spin dephasing parameter $\gm_2\equiv 1/T_2$,
and the resonator damping $\gm$. For magnetic nanomolecules, the characteristic
values are $\gm_1\sim(10^{-7}-10^{-5})$ s$^{-1}$ and $\gm_2\sim 10^{10}$
s$^{-1}$. Other attenuation mechanisms are discussed in Ref. [13].

The derivation of the equations of motion for the spin operators has been
described in detail earlier [3,11,13]. Therefore, here we present the final
form of these equations. To write down the equations in a compact way, we
introduce the notations
$$
\xi_i^0 \equiv \frac{1}{\hbar} \; \sum_{j(\neq i)}
\left ( a_{ij} S_j^z +  c_{ij}^* S_j^- + c_{ij} S_j^+ \right ) \; ,
$$
\be
\label{11}
\xi_i \equiv \frac{i}{\hbar} \; \sum_{j(\neq i)}
\left ( 2 c_{ij} S_j^z - \; \frac{1}{2} \; a_{ij} S_j^- +
2b_{ij} S_j^+ \right ) \; ,
\ee
in which $S_j^{\pm}$ are the ladder spin operators and the dipolar coefficients
are
\be
\label{12}
a_{ij} \equiv D_{ij}^{zz} \; , \qquad
b_{ij} \equiv \frac{1}{4} \left ( D_{ij}^{xx} - D_{ij}^{yy} -
2i D_{ij}^{xy} \right ) \; , \qquad
c_{ij} \equiv \frac{1}{2} \left ( D_{ij}^{xx} - i D_{ij}^{yz} \right ) \; .
\ee
Also, we define the effective force acting on  the $j$ spin as
\be
\label{13}
f_j \equiv - \; \frac{i}{\hbar} \; \mu_0 H + \xi_j \; .
\ee
Taking into account the saturation effect [13,28], the effective transverse
attenuation can be represented as
\be
\label{14}
\Gm_2 = \gm_2 \left ( 1 - s^2 \right ) \; ,
\ee
in which
\be
\label{15}
s \equiv \frac{1}{SN} \; \sum_{j=1}^N < S_j^z >
\ee
is the reduced longitudinal spin polarization. Using the above notations,
we obtain [3,11,13] the equations of motion for the ladder spin operator
$S_j^-\equiv S_j^x-iS_j^y$,
\be
\label{16}
\frac{dS_j^-}{dt} = - i \left ( \om_0 +\xi_j^0 -i\Gm_2 \right ) S_j^-
+ f_j S_j^z + i\; \frac{\om_D}{S} \; S_j^z S_j^- \; ,
\ee
and for the component $S_j^z$,
\be
\label{17}
\frac{dS_j^z}{dt} = - \; \frac{1}{2} \left ( f_j^+ S_j^- + S_j^+ f_j
\right ) - \gm_1 \left ( S_j^z - \zeta \right ) \; ,
\ee
where $\zeta$ is an equilibrium spin polarization.

Our aim is to study the behaviour of the reduced spin polarization (15)
as a function of time, $s=s(t)$, starting from a given initial polarization
$s_0\equiv s(0)$. The speed of the polarization reversal is connected with
the level of the magnetodipole radiation characterized by the radiation
intensity
\be
\label{18}
I(t) = \frac{2\mu^2_0}{3c^3} \left |
\sum_j < \ddot{\bS}_j > \right |^2 \; ,
\ee
where the overdots mean the time differentiation. In order to analyze to what
extent the radiation is coherent, we separate in the radiation intensity (18),
the coherent and incoherent parts,
\be
\label{19}
I(t) = I_{inc}(t) + I_{coh}(t) \; ,
\ee
with the incoherent and coherent radiation intensities
\be
\label{20}
I_{inc}(t) \equiv \frac{2\mu_0^2}{3c^3} \; \sum_j
\left | < \ddot{\bS}_j > \right |^2 \; , \qquad
I_{coh}(t) \equiv \frac{2\mu_0^2}{3c^3} \; \sum_{i\neq j}
 < \ddot{\bS}_i \ddot{\bS}_j >  \; ,
\ee
respectively.

We analyze the temporal evolution of $s(t)$ and $I(t)$ under different system
parameters. Our final goal is to find the conditions providing the largest
radiation intensity.

\section{Results of calculations}

We have investigated the equations of motion (16) and (17), with the
equation (6) for the resonator feedback field, in two ways. First, the
spins $S_j^\al$ have been treated as operator variables. Then, to find
the behaviour of the spin polarization (15), we need to average the
evolution equations (16) and (17). The following analysis is based on
the scale separation approach [2,3,6--9], which is an extension of the
averaging techniques [29] to the systems of stochastic differential
equations. The second way of analyzing Eqs. (16) and (17) is by computer
modeling, when $S_j^\al$ are treated as classical variables.

The first way, using the scale separation approach [2,3,6--9], better
describes the initial stage of spin relaxation, when the coherence of
spin motion has not yet been developed and quantum effects prevail. At
this stage, the radiation intensity is very low, since the spins move
chaotically. Such a chaotic stage lasts during the {\it chaos time} $t_c$,
which can be estimated [13] as $t_c\sim(1-s_0^2)/\om_0 s_0$. For $s_0=0.9$
and $\om_0\sim 10^{13}$ s$^{-1}$, this gives $t_c\sim 10^{-14}$ s. After the
time $t_c$, the dynamic coherence quickly develops, when the semiclassical
approximation becomes applicable. At this coherent stage, the direct computer
simulation gives a more appropriate description. The results of both methods
of calculations, as we have checked, are close to each other, provided that
in computer modeling one starts with $s_0<1$. The main difference is that
the scale separation approach, being a kind of the averaging techniques,
yields, for the solutions of the evolution equations their guiding centers,
with fast oscillations being smoothed out. Since, at the coherent stage, the
semiclassical approximation is well justified, we present below the results
of direct numerical modeling, employing Eqs. (16) and (17), where spins are
treated as classical variables.

In the figures below, time is measured in units of $T_2$ and all frequencies,
in units of $\gm_2$. We assume the resonance condition $\om_0=\om$. The
initial spin polarization in all figures, except Figs. 4 and 5, is taken as
$s_0=0.9$. Other system parameters are listed in the related figure captions.

The radiation intensity is shown in reduced units, so that its maximum be
close to one. In order to connect the calculated intensity with physical
units, we measure our numerical results using the units of $N^2I_0$, where
$N$ is the number of molecular spins and
$$
I_0 \equiv \frac{2\mu_0^2}{3c^3} \; \gm_2^4 \qquad (\mu_0=-2\mu_B)
$$
is a characteristic radiation intensity of a single spin. With
$\mu_B=0.9274\times 10^{-20}$ erg/G, $\mu_0=1.855\times 10^{-20}$ erg/G,
and $\gm_2=10^{10}$ s$^{-1}$, we have $I_0=0.852\times 10^{-38}$ W, where
1W$=10^7$ erg/s.

In Figs. 1 and 2, we study the influence of the Zeeman frequency $\om_0$
on the speed of polarization reversal and the level of radiation intensity.
The fastest reversal and the strongest radiation occurs for the largest
Zeeman frequency. Note that the present-day experimental facilities allow
one to reach quite large values of magnetic field [30] and, consequently,
to satisfy the inequality $\om_0\gg\om_D$, which is necessary for achieving
the resonance condition (10).

Figure 3 shows how the relaxation process depends on the value of the
resonator damping $\gm$. There exists an optimal value of $\gm$, for which
the final spin relaxation is kept fixed at the lowest level. However, for
achieving the fastest reversal and the strongest radiation, the smallest
$\gm$ is preferable.

The role of the initial spin polarization $s_0$ is demonstrated in Figs.
4 and 5. As is seen, the fastest polarization reversal and the strongest
radiation intensity happens under the largest initial polarization.

Figures 6 and 7 show how the anisotropy frequency $\om_D$ hinders the
resonance condition (10). The larger $\om_D$, the stronger it suppresses
the velocity of spin reversal and the maximal value of the radiation
intensity.

Dipole interactions also suppress the relaxation process and the radiation
intensity, as is seen in Figs. 8 and 9. The maximal intensity is diminished
by the factor of 1.52. However, the dipole interactions are inavoidable in
spin systems.

It is appropriate to recollect it again that dipole interactions do not
allow for the realization of spin superradiance in resonatorless magnets
[3,11--14,27]. Decoherent influence of dipole interactions can be overcome
only by coupling the spin sample with a resonator producing sufficiently
strong feedback field.

In addition to the necessary presence of a resonator, the role of dipole
interactions can be regulated by varying the sample shape and orientation
with respect to the external magnetic field and the direction of the
resonator feedback field. This is illustrated in Figs. 10 and 11 which show
that the best spin reversal and the strongest radiation intensity corresponds
to the case of a spin chain directed along the axis of the resonant magnetic
coil, that is, along the direction of the resonator feedback field.

To estimate the power of radiation intensity that can be obtained in real
physical systems, we may accept for the number of coherently radiating
molecules the coherence number
$$
N_{coh} = \rho \lbd^3 \qquad \left ( \lbd \equiv
\frac{2\pi c}{\om_0} \right ) \; .
$$
Taking the typical density of magnetic molecules in molecular magnets
$\rho\approx 0.4\times 10^{21}$ cm$^{-3}$ and setting $\om_0=2\times 10^{13}$
s$^{-1}$, with $\lbd=0.943\times 10^{-2}$ cm, we get $N_{coh}\approx 10^{14}$.
Then the maximal value of the radiation intensity (18) can be as high as
$I_{max}\sim 10^5$ W. Comparing the incoherent and coherent parts of the
intensity, as defined in Eq. (20), shows that the radiation in its maximum
is practically completely coherent.

In conclusion, we have demonstrated that nanomolecues, as well as
nanoclusters, can produce coherent spin radiation, provided they are coupled
to a resonant electric circuit. Spin reversal can be made very fast, of the
order of $10^{-11}$ s, which defines the duration of the coherent radiation
pulse $t_p\sim 10^{-11}$ s. For magnetic nanomolecues, the maximal radiation
intensity can reach $10^5$ W. To achieve the strongest radiation requires to
choose the largest $\om_0$ and $s_0$, but the smallest $\gm$ and $\om_D$. The
sample shape and its orientation are important for obtaining the strongest
radiation. The most favourable configuration corresponds to an elongated spin
sample placed along the direction of the resonator feedback field.

\vskip 5mm

{\bf Acknowledgement}

\vskip 2mm

Two of the authors (V.I.Y. and E.P.Y.) acknowledge financial support
from the Russian Foundation for Basic Research (Grant 08-02-00118).

\newpage

\begin{center}

{\bf{\Large Figure Captions} }
\end{center}

{\bf Fig. 1}. Reduced spin polarization $s(t)$ as a function of
dimensionless time (measured in units of $T_2$) for a cubic sample
of $N=125$ molecules with spin $S=10$. Initial reduced polarization
is $s_0=0.9$, the anisotropy frequency is $\om_D=20$, and the resonator
damping is $\gm=10$. The Zeeman frequency is varied: $\om_0=1000$ (solid
line),  $\om_0=2000$ (long-dashed line), and $\om_0=5000$ (short-dashed
line). The fastest polarization reversal is for the largest Zeeman
frequency.

\vskip 5mm

{\bf Fig. 2}. Radiation intensity $I(t)$ for the parameters of Fig. 1
for different Zeeman frequencies: $\om_0=1000$ (solid line), intensity
is in units of $3.199\times 10^{13}$ N$^2$I$_0$; $\om_0=2000$ (long-dashed
line), intensity is in units of $0.832\times 10^{15}$ N$^2$I$_0$;
$\om_0=5000$ (short-dashed line), intensity is in units of $5.118\times
10^{16}$ N$^2$I$_0$. The strongest intensity is for the largest Zeeman
frequency.

\vskip 5mm

{\bf Fig. 3} Reduced spin polarization $s(t)$ as a function of
dimensionless time for a cubic sample of  N=125 molecules, with spin
$S=10$, initial polarization $s_0=0.9$, Zeeman frequency $\om_0=2000$,
and anisotropy frequency $\om_D=20$, for different resonator dampings:
$\gm=1$ (solid line), $\gm=10$ (long-dashed line), and $\gm=50$
(short-dashed line). The fastest polarization reversal is for the
smallest resonator damping, which yields the largest radiation intensity
of order $10^{15}$ N$^2$I$_0$.

\vskip 5mm

{\bf Fig. 4} Reduced spin polarization $s(t)$ as a function of
dimensionless time for a cubic sample of  $N=125$ molecules, with spin
$S=10$, Zeeman frequency $\om_0=2000$, anisotropy frequency $\om_D=20$,
resonator damping $\gm=10$, for different initial polarizations:
$s_0=0.9$ (solid line), $s_0=0.7$ (long-dashed line), and  $s_0=0.5$
(short-dashed line). The fastest polarization reversal occurs for the
largest initial polarization.

\vskip 5mm

{\bf Fig. 5}. Radiation intensity $I(t)$ for the parameters of Fig. 4
for varying initial polarizations: $s_0=0.9$ (solid line),  $s_0=0.7$
(long-dashed line), and $s_0=0.5$ (short-dashed line). Intensity units
are $0.832\times 10^{15}$ N$^2$I$_0$. The strongest radiation intensity
is for the largest initial polarization.

\vskip 5mm

{\bf Fig. 6}. Reduced spin polarization $s(t)$ as a function of
dimensionless time for a cubic sample of $N=125$ molecules of spin
$S=10$, with the Zeeman frequency $\om_0=2000$ and resonator damping
$\gm=10$, for different magnetic anisotropy frequencies: $\om_D=20$
(solid line), $\om_D=50$ (long-dashed line), and $\om_D=100$ (short-dashed
line). The fastest polarization reversal occurs for the smallest magnetic
anisotropy.

\vskip 5mm

{\bf Fig. 7}. Radiation intensity $I(t)$ for the parameters of Fig. 6
for varying magnetic anisotropy frequencies: $\om_D=20$ (solid line),
$\om_D=50$ (long-dashed line), and $\om_D=100$ (short-dashed line). The
values of the radiation intensity are in units of $0.832\times 10^{15}$
N$^2$I$_0$. The strongest radiation  intensity is for the weakest
magnetic anisotropy.

\vskip 5mm

{\bf Fig. 8}. Reduced spin polarization $s(t)$ as a function of
dimensionless time for a cubic sample of $N=125$ molecules of spin
$S=10$, with $\om_0=2000$, $\om_D=20$, and $\gm=10$, for the case
of present dipole interactions (solid line) and the case when they
are absent (dashed line). The polarization reversal is hindered by
dipole interactions.

\vskip 5mm

{\bf Fig. 9}. Radiation intensity $I(t)$ for the parameters of Fig. 8
for the cases with dipole interactions (solid line) and without these
interactions (dashed line). Solid line corresponds to the intensity
in units $0.806\times 10^{15}$ N$^2$I$_0$ and dashed line, in units
$1.228\times 10^{15}$ N$^2$I$_0$. Radiation intensity is supressed by
dipole interactions by the factor 1.524.

\vskip 5mm

{\bf Fig. 10}. Reduced spin polarization $s(t)$ as a function
of dimensionless time for $N=144$ molecules of spin $S=10$, with
$\om_0=2000$, $\om_D=20$, and $\gm=30$, for different sample shapes
and orientations: the chain of spins along the $z$-axis (solid line),
the chain along the $x$-axis (long-dashed line), the $y-z$ plane of
spins (short-dashed line), and the $x-y$ plane of spins (dotted-dashed
line). The fastest and strongest spin reversal occurs for the case of
the spin chain along the $x$-axis coinciding with the resonator axis.

\vskip 5mm

{\bf Fig. 11}. Radiation intensity $I(t)$ as a function of dimensionless
time for $N=144$ molecules of spin $S=10$, with the same parameters as in
Fig. 10, for different sample shapes and orientations: the spin chain along
the $z$-axis (solid line), the spin chain along the $x$-axis (long-dashed
line), the $y-z$ spin plane (short-dashed line), and the $x-y$ spin plane
(dashed-dotted line). The strongest radiation intensity is for the spin
chain along the $x$-axis, reaching in the maximum $1.206\times 10^{15}$
N$^2$I$_0$.

\newpage

\begin{figure}[ht]
\centerline{\psfig{file=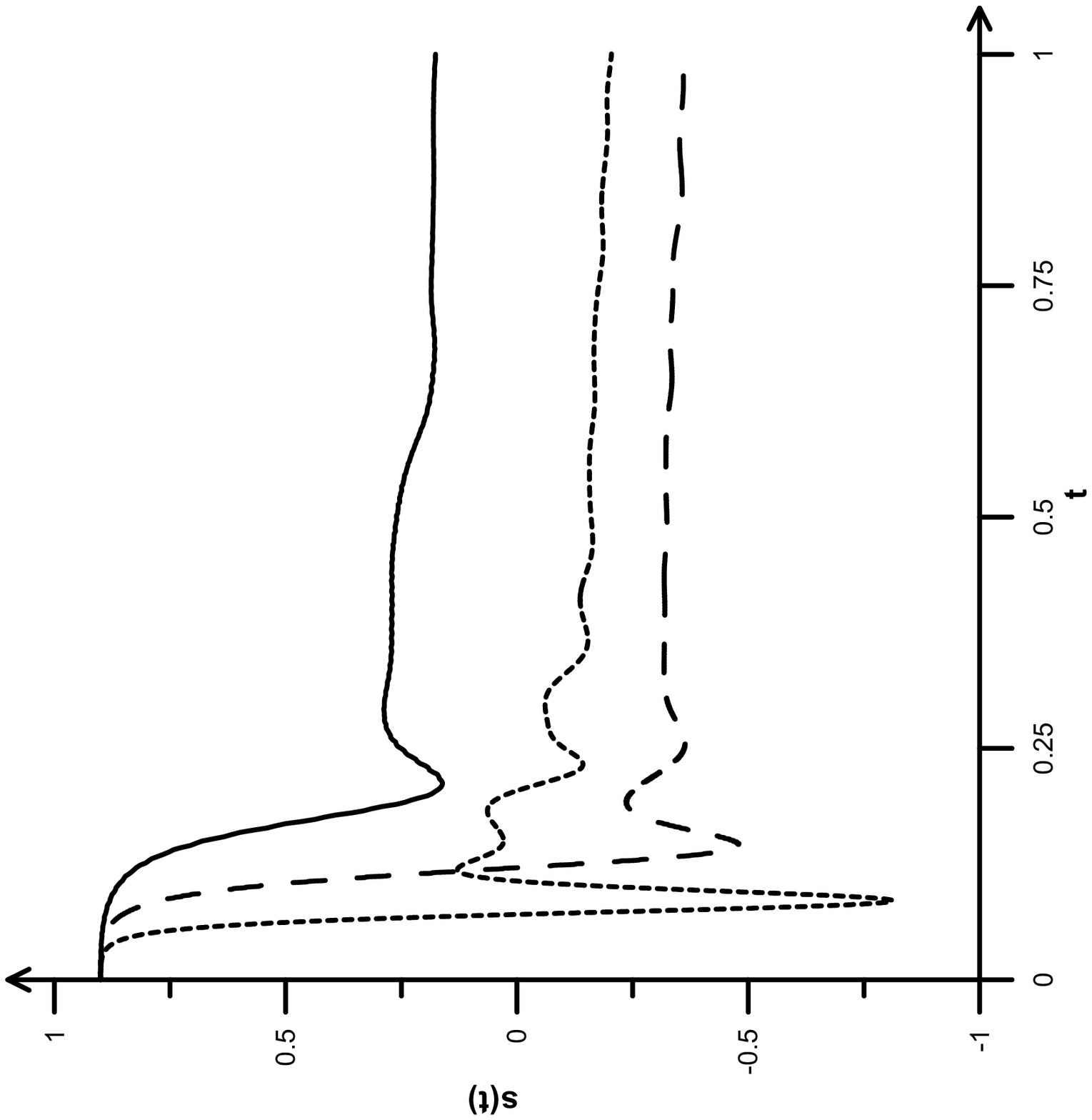,angle=0,width=10cm}}
\caption{Reduced spin polarization $s(t)$ as a function of
dimensionless time (measured in units of $T_2$) for a cubic sample
of $N=125$ molecules with spin $S=10$. Initial reduced polarization
is $s_0=0.9$, the anisotropy frequency is $\om_D=20$, and the resonator
damping is $\gm=10$. The Zeeman frequency is varied: $\om_0=1000$ (solid
line),  $\om_0=2000$ (long-dashed line), and $\om_0=5000$ (short-dashed
line). The fastest polarization reversal is for the largest Zeeman
frequency.
}
\label{fig:Fig.1}
\end{figure}

\newpage

\begin{figure}[hbtp]
\vspace{9pt}
\centerline{
\hbox{
\includegraphics[width=7cm]{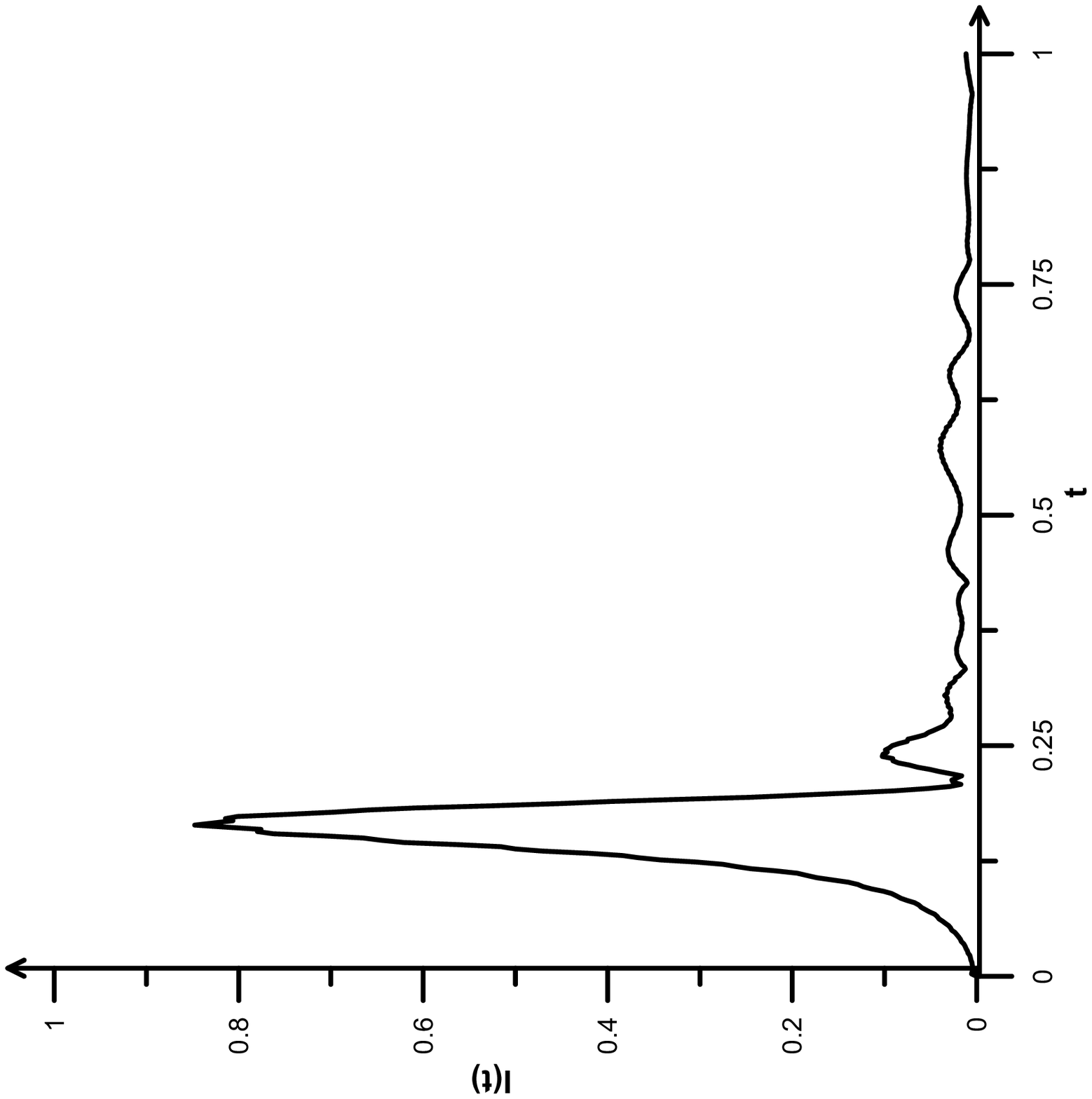} \hspace{2cm}
\includegraphics[width=7cm]{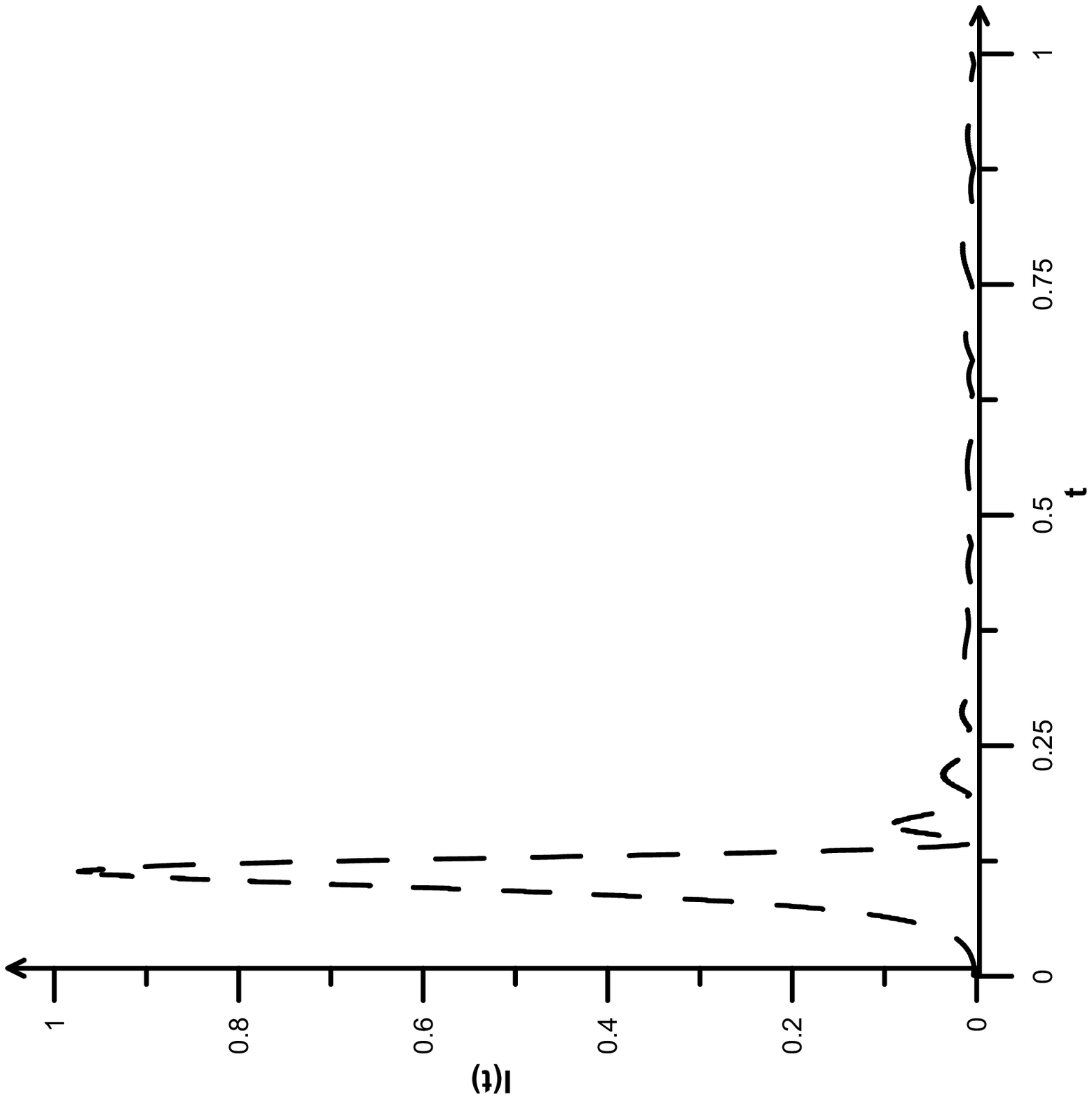} }}
\vspace{9pt}
\centerline{
\hbox{
\includegraphics[width=7cm]{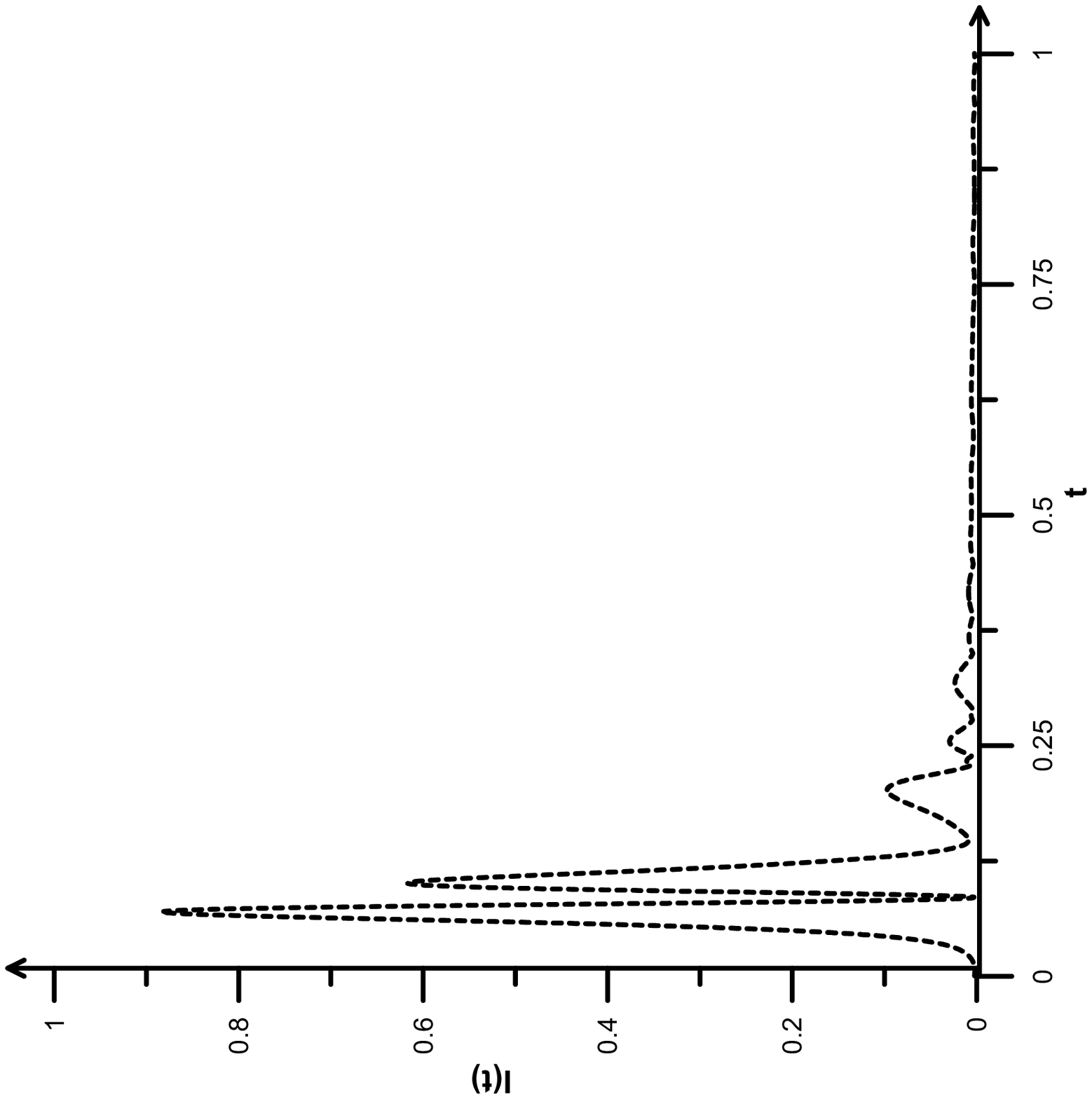} \hspace{2cm}
}}
\caption{Radiation intensity $I(t)$ for the parameters of Fig. 1 for
different Zeeman frequencies: $\om_0=1000$ (solid line), intensity is
in units of $3.199\times 10^{13}$ N$^2$I$_0$; $\om_0=2000$ (long-dashed
line), intensity is in units of $0.832\times 10^{15}$ N$^2$I$_0$;
$\om_0=5000$ (short-dashed line), intensity is in units of $5.118\times
10^{16}$ N$^2$I$_0$. The strongest intensity is for the largest Zeeman
frequency.}
\label{fig:Fig.2}
\end{figure}

\newpage

\begin{figure}[ht]
\centerline{\psfig{file=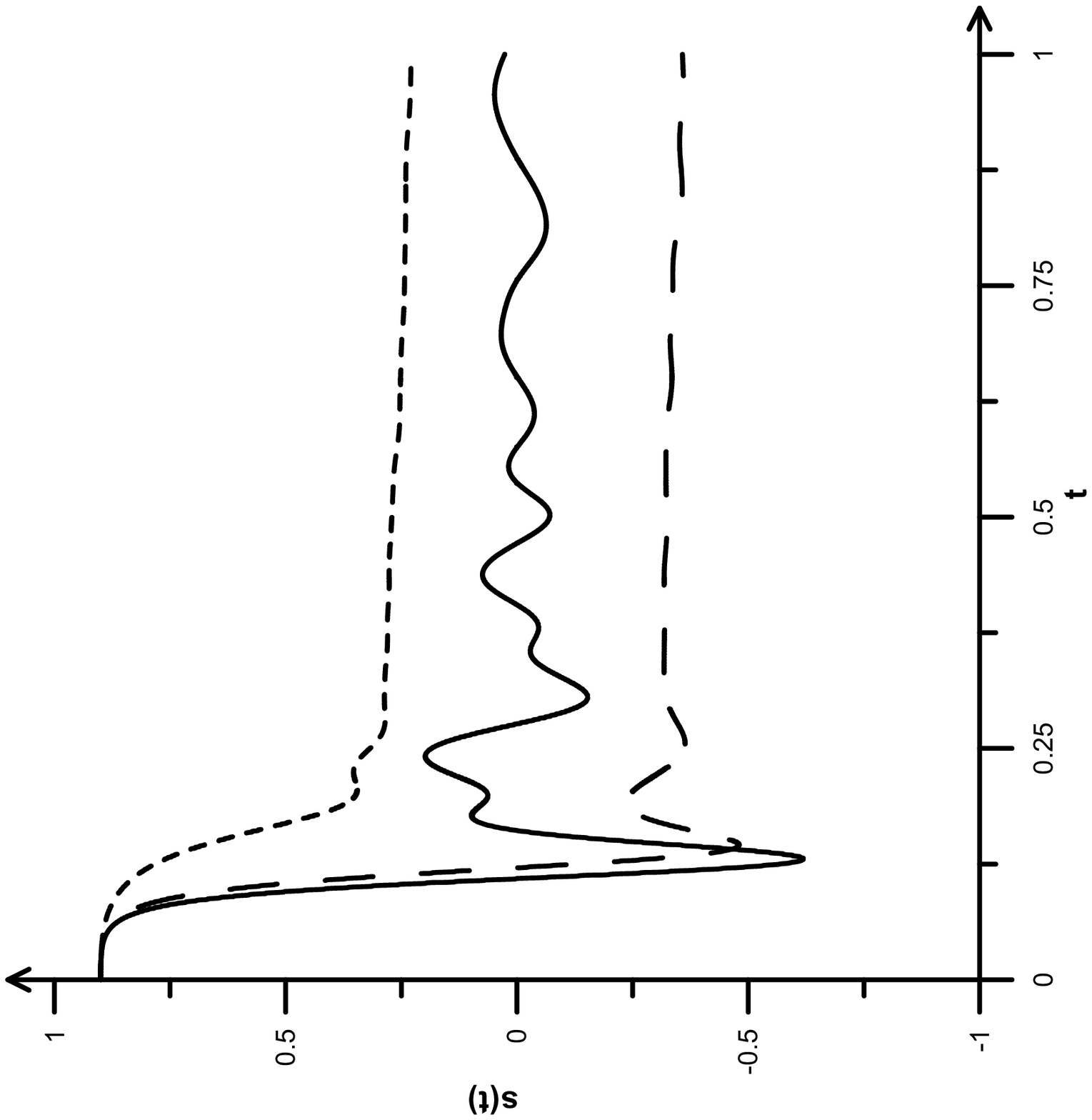,angle=0,width=10cm}}
\caption{Reduced spin polarization $s(t)$ as a function of
dimensionless time for a cubic sample of  N=125 molecules, with spin
$S=10$, initial polarization $s_0=0.9$, Zeeman frequency $\om_0=2000$,
and anisotropy frequency $\om_D=20$, for different resonator dampings:
$\gm=1$ (solid line), $\gm=10$ (long-dashed line), and $\gm=50$
(short-dashed line). The fastest polarization reversal is for the
smallest resonator damping, which yields the largest radiation intensity
of order $10^{15}$ N$^2$I$_0$.
}
\label{fig:Fig.3}
\end{figure}

\begin{figure}[ht]
\centerline{\psfig{file=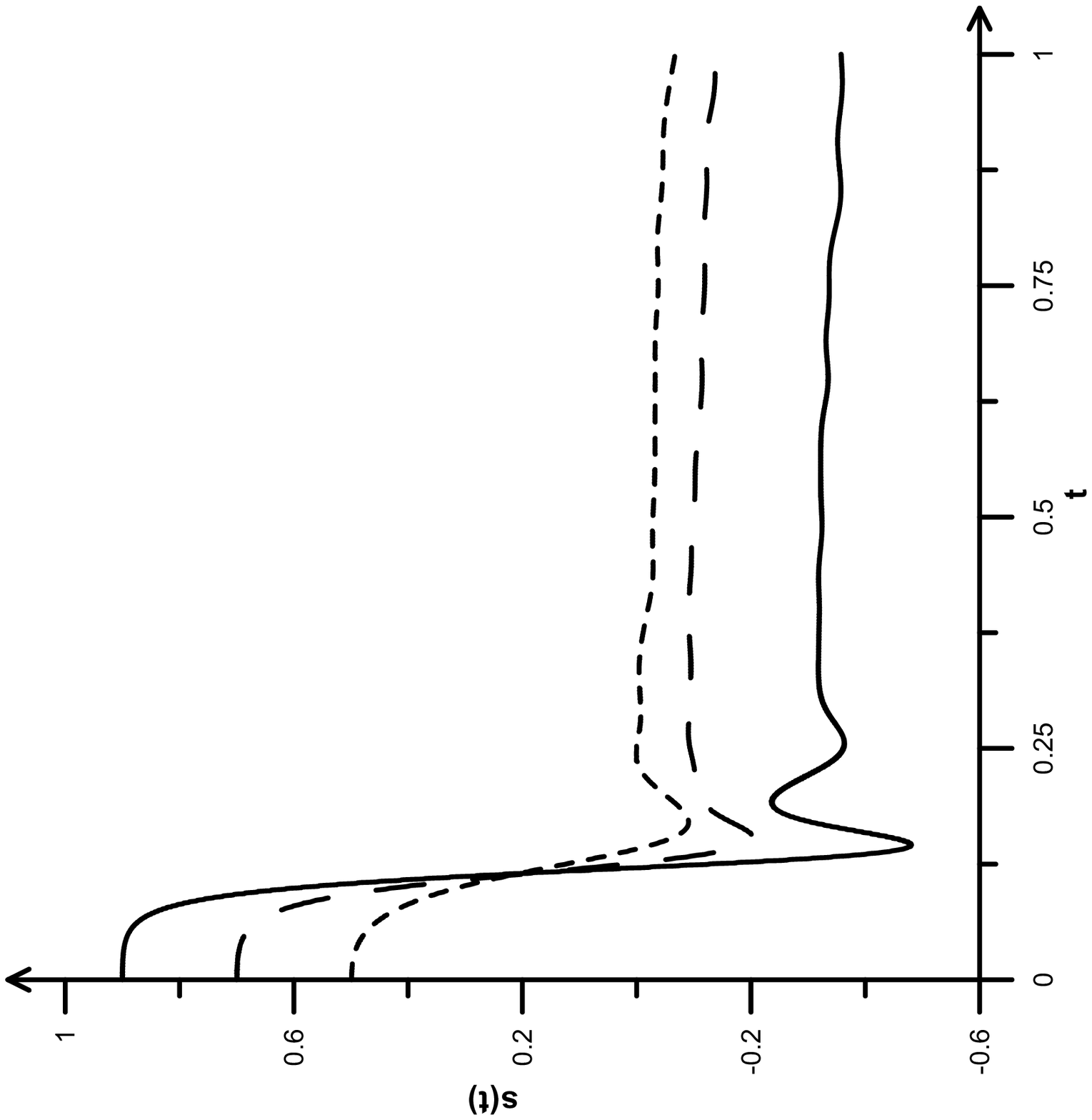,angle=0,width=10cm}}
\caption{Reduced spin polarization $s(t)$ as a function of
dimensionless time for a cubic sample of  $N=125$ molecules, with spin
$S=10$, Zeeman frequency $\om_0=2000$, anisotropy frequency $\om_D=20$,
resonator damping $\gm=10$, for different initial polarizations:
$s_0=0.9$ (solid line), $s_0=0.7$ (long-dashed line), and  $s_0=0.5$
(short-dashed line). The fastest polarization reversal occurs for the
largest initial polarization.
}
\label{fig:Fig.4}
\end{figure}

\newpage

\begin{figure}[ht]
\centerline{\psfig{file=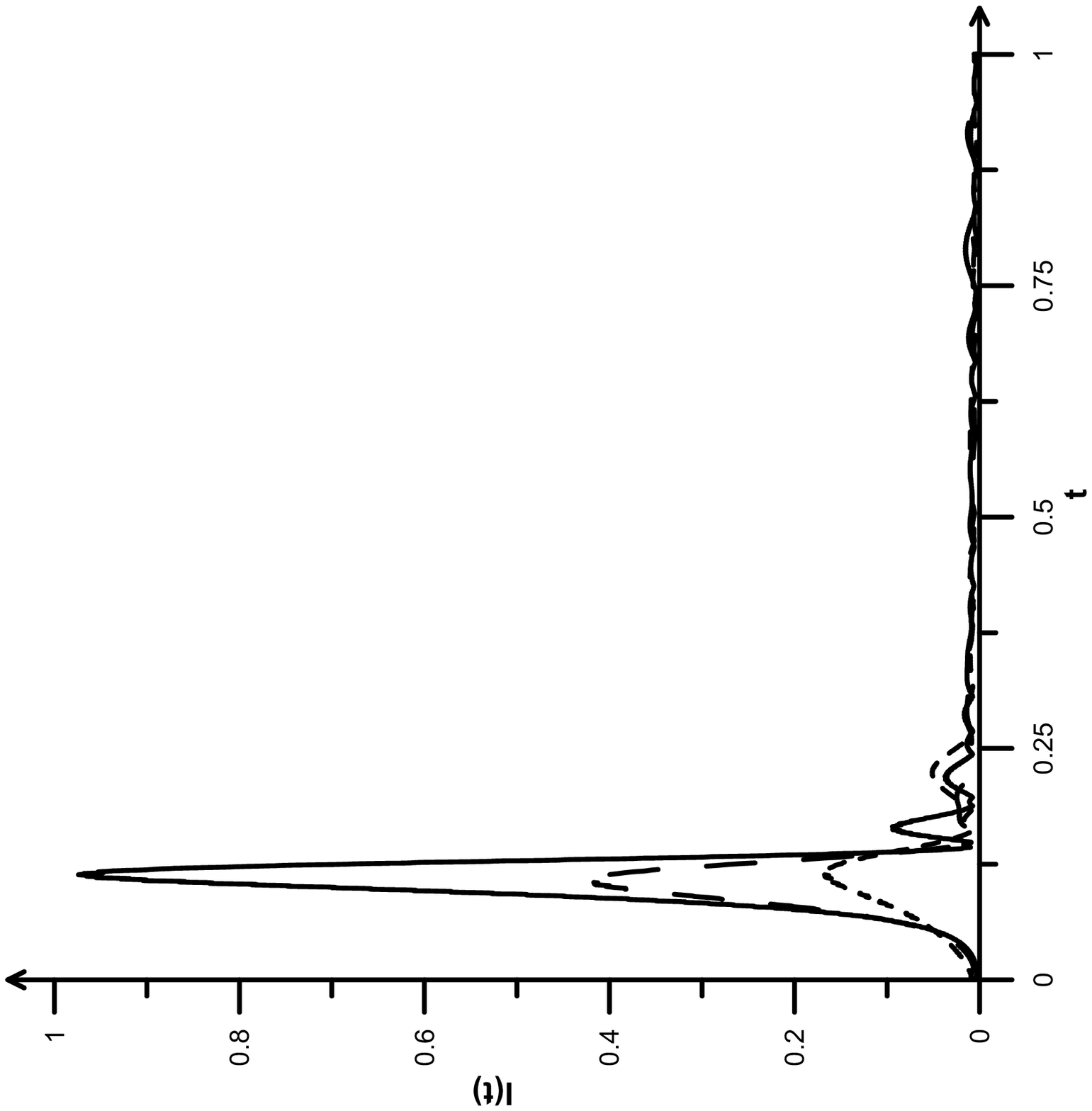,angle=0,width=10cm}}
\caption{Radiation intensity $I(t)$ for the parameters of Fig. 4
for varying initial polarizations: $s_0=0.9$ (solid line),  $s_0=0.7$
(long-dashed line), and $s_0=0.5$ (short-dashed line). Intensity units
are $0.832\times 10^{15}$ N$^2$I$_0$. The strongest radiation intensity
is for the largest initial polarization.
}
\label{fig:Fig.5}
\end{figure}

\begin{figure}[ht]
\centerline{\psfig{file=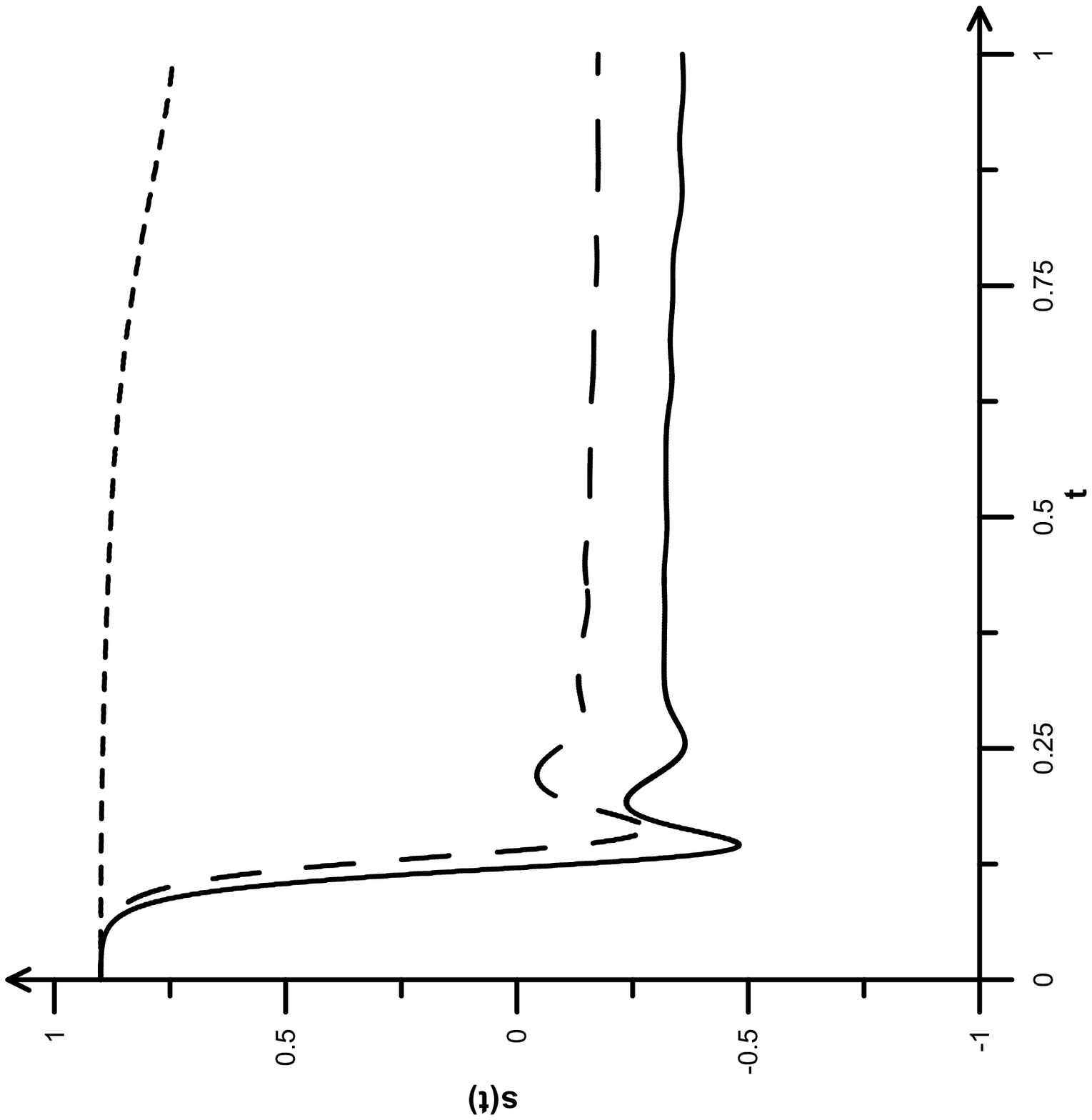,angle=0,width=10cm}}
\caption{Reduced spin polarization $s(t)$ as a function of
dimensionless time for a cubic sample of $N=125$ molecules of spin
$S=10$, with the Zeeman frequency $\om_0=2000$ and resonator damping
$\gm=10$, for different magnetic anisotropy frequencies: $\om_D=20$
(solid line), $\om_D=50$ (long-dashed line), and $\om_D=100$ (short-dashed
line). The fastest polarization reversal occurs for the smallest magnetic
anisotropy.
}
\label{fig:Fig.6}
\end{figure}

\newpage

\begin{figure}[ht]
\centerline{\psfig{file=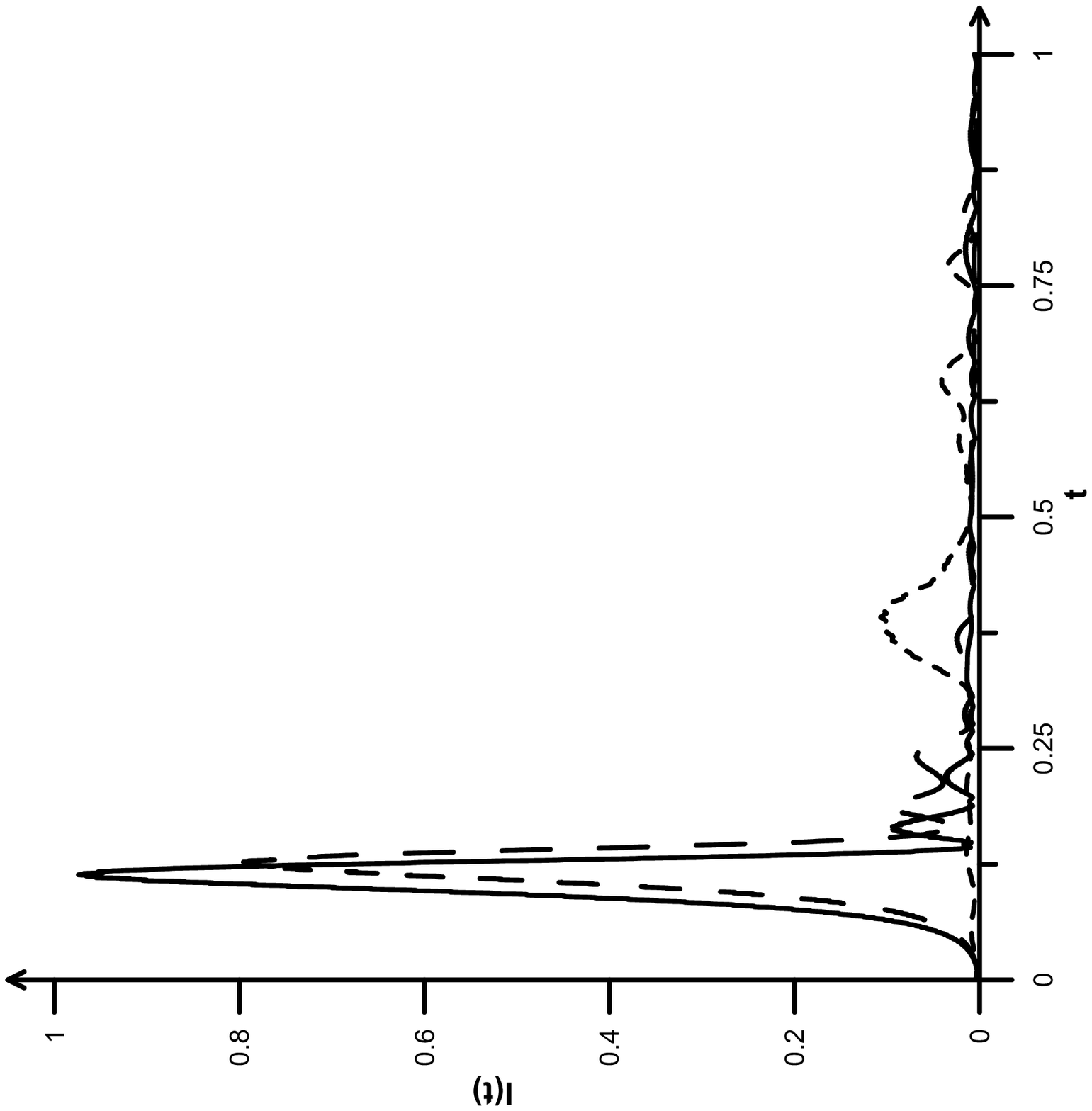,angle=0,width=10cm}}
\caption{Radiation intensity $I(t)$ for the parameters of Fig. 6
for varying magnetic anisotropy frequencies: $\om_D=20$ (solid line),
$\om_D=50$ (long-dashed line), and $\om_D=100$ (short-dashed line). The
values of the radiation intensity are in units of $0.832\times 10^{15}$
N$^2$I$_0$. The strongest radiation  intensity is for the weakest
magnetic anisotropy.
}
\label{fig:Fig.7}
\end{figure}

\newpage

\begin{figure}[ht]
\centerline{\psfig{file=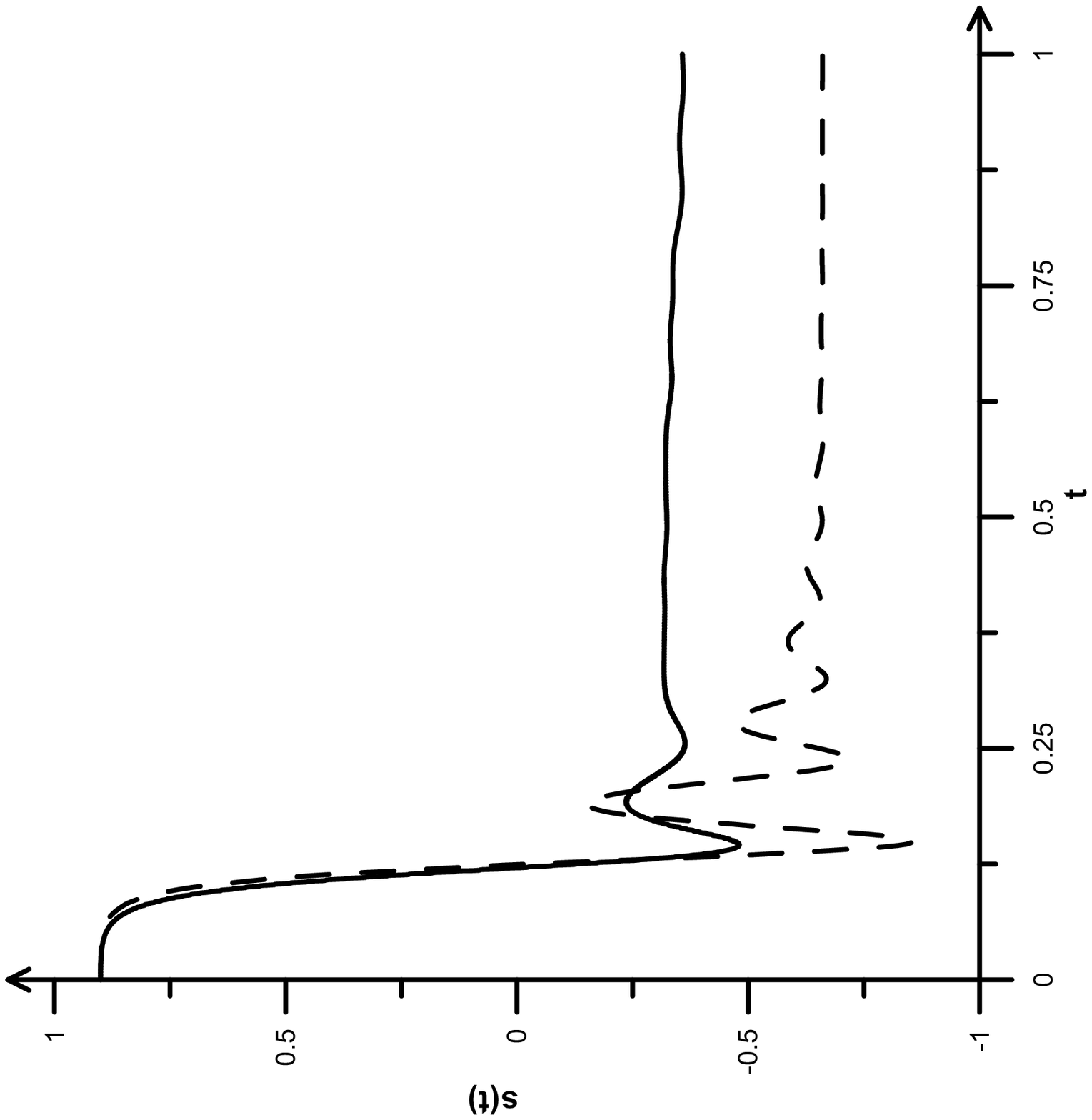,angle=0,width=10cm}}
\caption{Reduced spin polarization $s(t)$ as a function of
dimensionless time for a cubic sample of $N=125$ molecules of spin
$S=10$, with $\om_0=2000$, $\om_D=20$, and $\gm=10$, for the case
of present dipole interactions (solid line) and the case when they
are absent (dashed line). The polarization reversal is hindered by
dipole interactions.
}
\label{fig:Fig.8}
\end{figure}

\begin{figure}[ht]
\centerline{\psfig{file=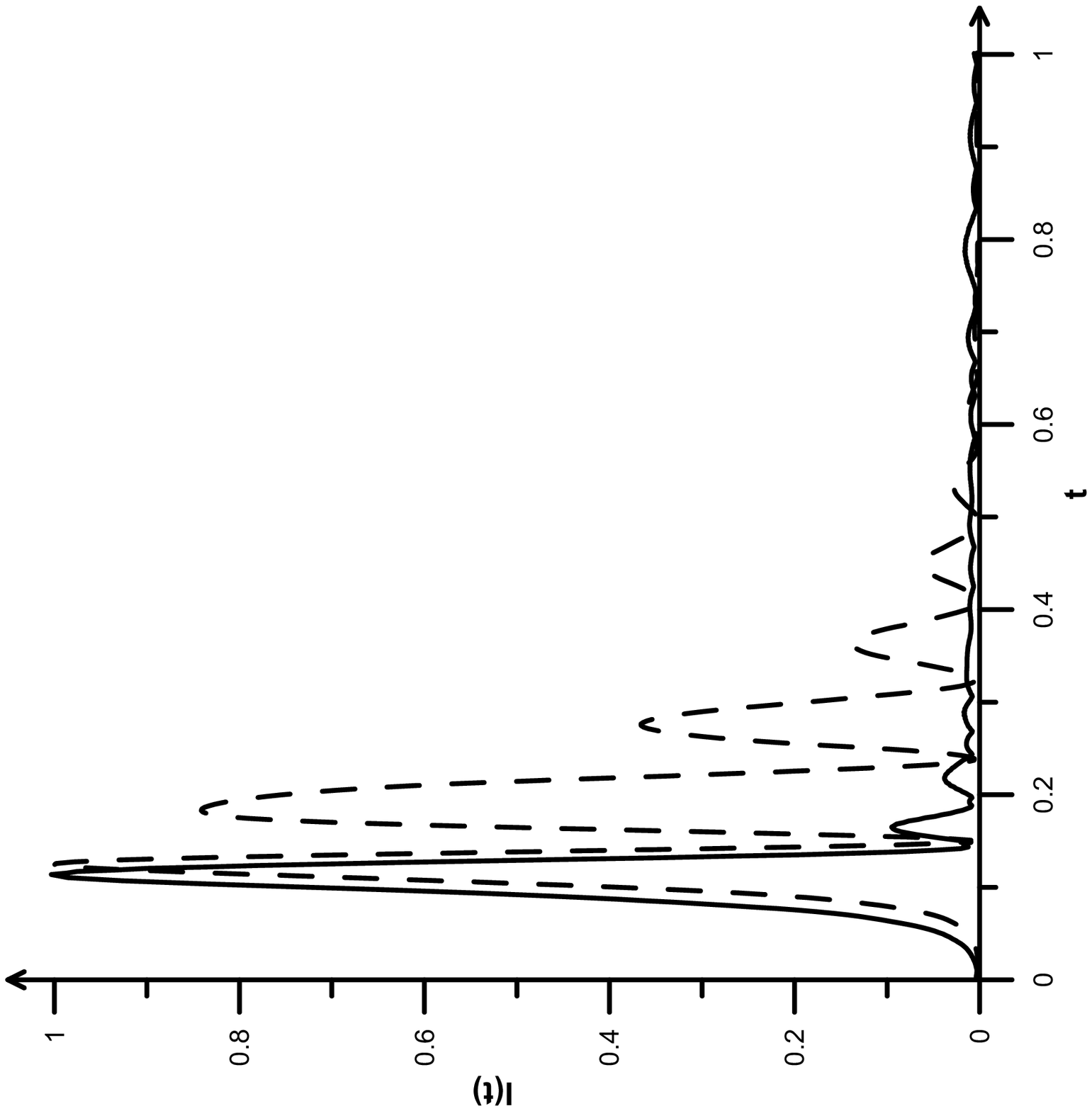,angle=0,width=10cm}}
\caption{Radiation intensity $I(t)$ for the parameters of Fig. 8
for the cases with dipole interactions (solid line) and without these
interactions (dashed line). Solid line corresponds to the intensity
in units $0.806\times 10^{15}$ N$^2$I$_0$ and dashed line, in units
$1.228\times 10^{15}$ N$^2$I$_0$. Radiation intensity is supressed by
dipole interactions by the factor 1.524.
}
\label{fig:Fig.9}
\end{figure}

\newpage

\begin{figure}[ht]
\centerline{\psfig{file=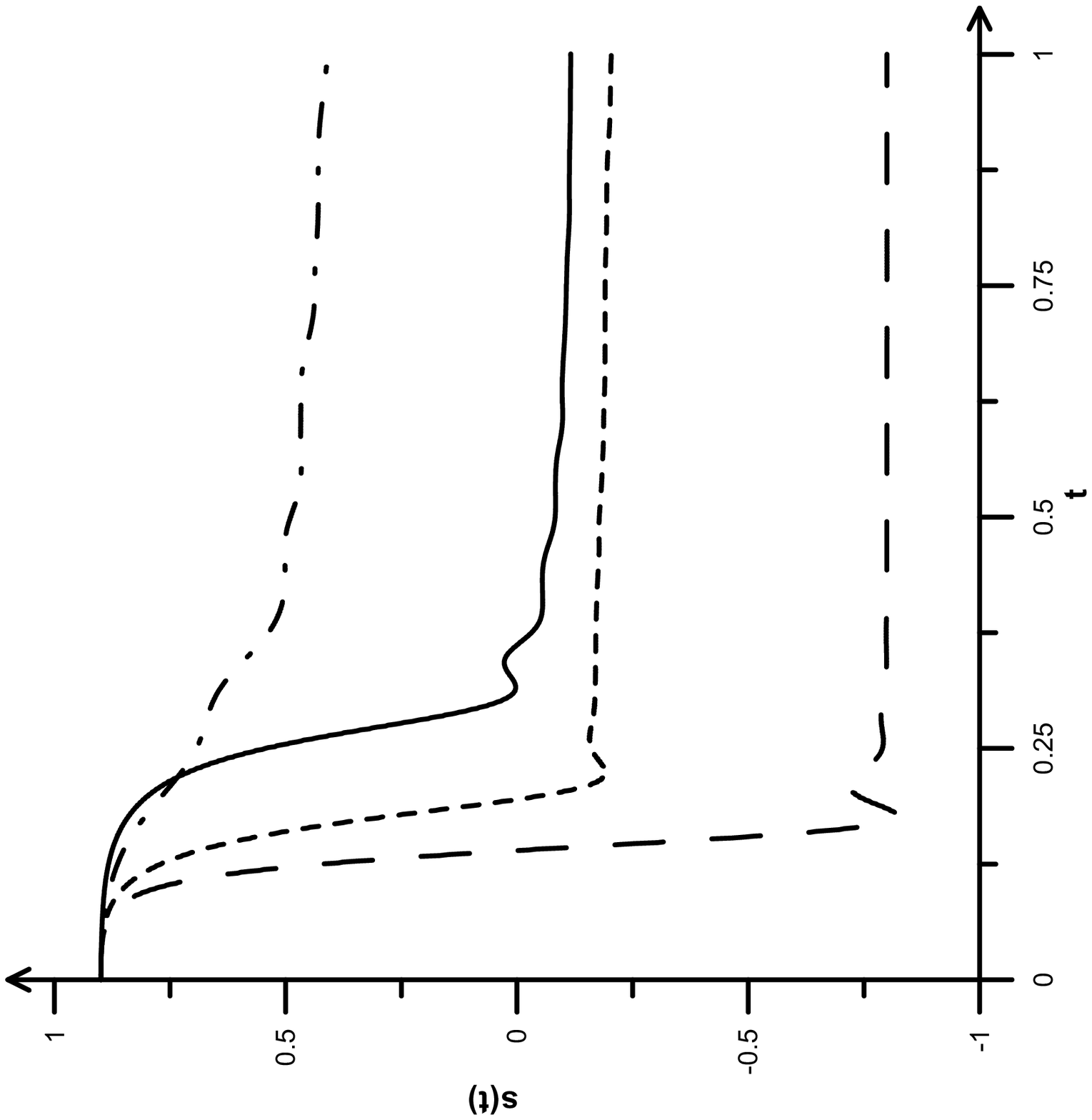,angle=0,width=10cm}}
\caption{Reduced spin polarization $s(t)$ as a function
of dimensionless time for $N=144$ molecules of spin $S=10$, with
$\om_0=2000$, $\om_D=20$, and $\gm=30$, for different sample shapes
and orientations: the chain of spins along the $z$-axis (solid line),
the chain along the $x$-axis (long-dashed line), the $y-z$ plane of
spins (short-dashed line), and the $x-y$ plane of spins (dotted-dashed
line). The fastest and strongest spin reversal occurs for the case of
the spin chain along the $x$-axis coinciding with the resonator axis.
}
\label{fig:Fig.10}
\end{figure}

\newpage

\begin{figure}[ht]
\centerline{\psfig{file=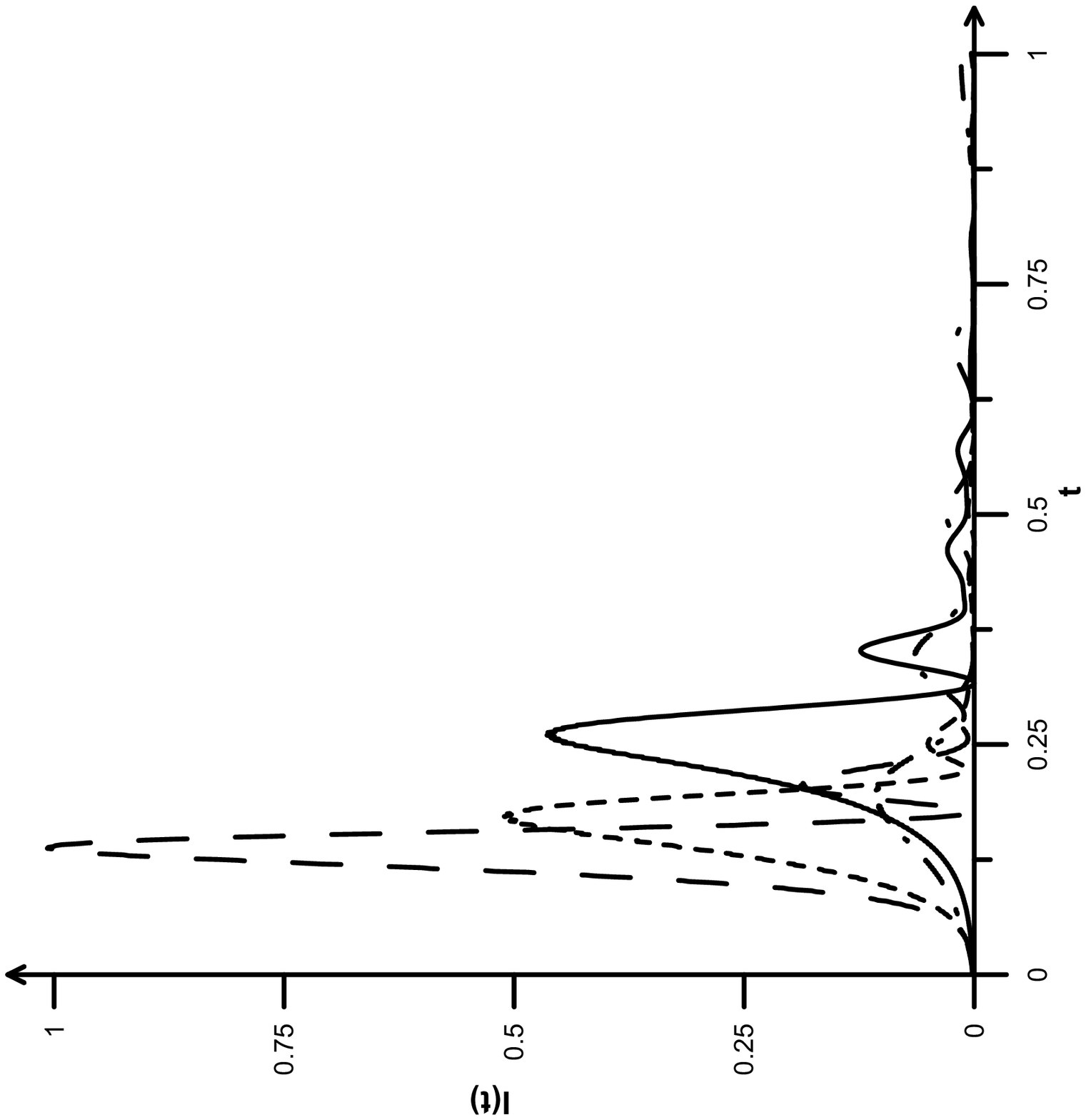,angle=0,width=10cm}}
\caption{Radiation intensity $I(t)$ as a function of dimensionless
time for $N=144$ molecules of spin $S=10$, with the same parameters as in
Fig. 10, for different sample shapes and orientations: the spin chain along
the $z$-axis (solid line), the spin chain along the $x$-axis (long-dashed
line), the $y-z$ spin plane (short-dashed line), and the $x-y$ spin plane
(dashed-dotted line). The strongest radiation intensity is for the spin
chain along the $x$-axis, reaching in the maximum $1.206\times 10^{15}$
N$^2$I$_0$.
}
\label{fig:Fig.11}
\end{figure}

\end{document}